\documentstyle[editedvolume,psfig]{crckapb}

\newcommand {\asec} {$^{\prime\prime}$}
\newcommand {\ti} {T$_{int}$~}
\newcommand {\tsa} {{\em tsa}~}
\newcommand {\cad} {{\em calibration darks}}
\newcommand {\ded} {{\em deactivation darks}}
\newcommand {\pad} {{\em parallel-mode darks}}
\newcommand {\hod} {{\em handover darks}}

\begin{opening}
\title{THE ISOCAM/LW DETECTOR DARK CURRENT BEHAVIOUR}

\author{A. BIVIANO}
\institute{Osservatorio Astronomico di Trieste, \\
via G.B. Tiepolo 11, I-34131 Trieste, Italy}
\institute{ISO Science Operation Centre, \\
Astrophysics Division, Space Science Department of ESA, \\
Villafranca, P.O. Box 50727, 28080 Madrid, Spain}
\author{M. SAUVAGE, P. GALLAIS, O. BOULADE}
\institute{CEA, DSM/DAPNIA/SAp, \\
CE-Saclay, F-91191, Gif-sur-Yvette Cedex, France}
\author{P. ROMAN, S. GUEST, K. OKUMURA, S. OTT}
\institute{ISO Science Operation Centre, \\
Astrophysics Division, Space Science Department of ESA, \\
Villafranca, P.O. Box 50727, 28080 Madrid, Spain}

\end{opening}

\runningtitle{ISOCAM DARK CURRENT}

\begin{document}

\begin{abstract}
We describe the calibration, measurements and data reduction, of the
dark current of the ISOCAM/LW detector.  We point-out the existence
of two significant drifts of the LW dark-current, one throughout the
ISO mission, on a timescale of days, another within each single
revolution, on a timescale of hours. We also show the existence of a
dependence of the dark current on the temperature of the ISOCAM detector.

By characterizing all these effects through polynomial fittings, we
build a model for the LW calibration dark, that depends on the epoch
of observation (parametrized with the revolution number and the time
elapsed in that given revolution since the activation) and on the
temperature of the ISOCAM detector. The model parameters are tuned for
each of ISOCAM/LW pixel.

We show that the modelling is very effective in taking into account
the dark-current variations and allows a much cleaner dark subtraction
than using a brute average of several calibration dark images.

The residuals of the LW model-dark subtraction are, on average,
similar to the pre-launch expectation.
\end{abstract}

\section{Introduction}\label{s-intro}
In this document we describe the calibration of the dark current in the
ISOCAM/LW detector. 

\begin{figure}
\centerline{\psfig{figure=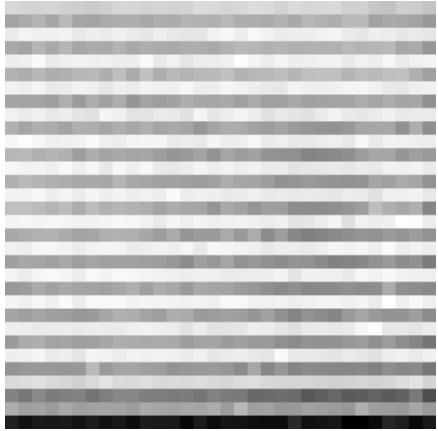,width=8cm}}
\caption[ ]{\footnotesize{An example of the LW dark frame, measured
at \ti=2~sec. Note the very evident line pattern.}}
\label{f-darkpat}
\end{figure}

The term ``dark current'' will be used throughout this paper to
indicate the level of the signal measured when the detector is in
darkness, i.e.  when no external flux reaches the detector.  Strictly
speaking, this is not a current, but an electronic reference level
including both real dark current and electrical offsets.  For the LW
detector, this signal is due to thermal charge generation in the
photoconductors and charge leakage generated during the commutation of
the reset transistor (see the ``Observer's Manual for ISOCAM").  A typical
example of the pattern measured on the LW detector in darkness is
shown in Fig.~\ref{f-darkpat} (for an integration time -- \ti hereafter --
of 2~sec). The LW dark frame shows a line pattern with a clear
separation between odd and even lines.

The output of the pixels of the detector at dark is given in the
standard units of all observations done with the ISOCAM/LW detector,
i.e. the {\em Analog-to-Digital Units} (ADU, hereafter). For a gain
factor $\times 2$, one ADU corresponds to 120 electrons above a given
electrical offset (see the ``Observer's Manual for ISOCAM'').  Since the
dark measurements are sometimes done with a gain factor $\times 1$,
and sometimes with a gain factor $\times 2$, for the sake of
simplicity in this document we scaled all units by the gain factor --
i.e. we use "ADU" to mean ``Analog-to-Digital Units scaled by the gain
factor''.

The ISOCAM dark measurements have been done at several integration
times, because the dark current (i.e. the measured output per pixel of
the detector in the darkness) has no simple linear dependence on the
integration time.  The pixel values can therefore be expressed in
terms of the ADU received per given integration time, or in terms of
the ADU received per second, by rescaling the measured output value by
the integration time. Hereafter, we will express the dark current
values either in ADU or in ADU/s, and specify the integration time
used in order to allow translation from one unit to another.

In summary, the observed value of the dark current, in ADU, on a given
pixel, is given by:
\begin{equation}
D_O = (N_{e^-} - offset)/(gain \times 60)
\label{e-darko}
\end{equation}
where $N_{e^-}$ is the number of electrons on the capacity of the
detector.

During the ISO mission it was shown that the LW dark-current is drifting both
on a timescale of hours within a given revolution, and on a timescale of days
throughout the mission. Hereafter we will refer to these drifts as the
{\em short-term} and {\em long-term} trends of the dark current.

The dark currents have been calibrated with four different kinds of
observations:
\begin{itemize}
\item Darks measured during calibration-dedicated revolutions,
i.e. those that are specifically dedicated to calibration observations
(typically, one per week); we call these measurements the
{\em calibration darks.}
\item Darks measured during the handover period, i.e. 
when the satellite tracking switches from one of the two antennae (VILSPA or
Goldstone) to the other (this time typically lasts 15--30~min every
revolution); we call these measurements the {\em handover darks.}
\item Darks measured during the deactivation period (lasting typically 1~hour
every revolution) i.e. when ISO enters the Earth radiation belts and
scientific observations are no longer possible because of the high rate of
glitch impacts, just before all instruments have to be switched off;
we call these measurements the {\em deactivation darks.}
\item Darks measured in the ISOCAM parallel mode of observation, i.e. taking
advantage of the part of telemetry (1/12) that is reserved to ISOCAM when
other ISO instruments are observing in prime mode; we call these measurements
the {\em parallel-mode darks.}
\end{itemize}

The four types of dark measurements have different goals. The 
\cad~ are long measurements done throughout the mission to constrain
the pattern of the dark frame, i.e. the dark current variation from one pixel
to another. The \cad~ are measured at regularly spaced time intervals,
typically one measure at all \ti every month. These time intervals are
not so dense as to allow a very clean characterization of the long-term
drifts of the dark currents. On the other hand, specific repeated observations
of the dark current throughout the same revolution allow to constrain the
short-term drift (as we will see in \S~\ref{s-short}).

A proper characterization of the long-term drifts is provided by the
\hod, which are shorter (and therefore noisier) measurements done at
every revolution, throughout the whole ISO mission, and more or less
at the same time after the activation of the instrument.
Unfortunately, the limited time available during the handover sequence
imposed severe constraints on the measurements, and it was decided to
monitor only two \ti (2 and 5~sec) LW darks during handover. Starting with
the revolution 764 (december 1997), the handover
sequence of dark measurements has been changed in order to monitor
also the 0.28 and 10~sec LW darks, as it was understood that the
long-term drift was important, and it was necessary to check that the
drift was the same (or at least similar) at different \ti (see
\S~\ref{s-long}). Since revolution 764, the even days were dedicated to
the measurements of darks at \ti=0.28 and 10~sec, and the odd days to
the darks at \ti=2 and 10~sec. The dark at \ti=20~sec was not covered,
because of time limitation (notice that ISOCAM observations
at \ti=20~sec are strongly affected from glitches, so that this \ti was
very seldom used by observers).

The \ded~ include LW dark measurements at \ti=2 and 5~sec.  One could
think that the \ded~ provide a monitoring of the long-term trend of
the dark currents. Unfortunately, these \ded~ are strongly affected by
the so-called {\em space weather,} i.e.  particles in the Earth
environment (for further details, see Gallais \& Boulade 1998).

Most of the measurements of \pad~ were done in the last seven months
of the ISO mission. Based on the repeated measurements of \cad~ in a
single revolution, the short-term drift of the LW dark current was
discovered. As a consequence, a denser time monitoring of the LW dark
current was required.  In order to have more measurements, and to
avoid using too much calibration time, it was decided to have dark
measurements with ISOCAM in the parallel mode.  Even with the reduced
telemetry rate available, these measurements have since proven to be
extremely useful to constrain the dark drifts within single
revolutions. These measurements have been done at all LW \ti, with the
exception of \ti=0.28~sec, unfeasible because of a software limitation
in the total number of frames that can be accumulated on-board.
However, several \cad~ have been measured at this particular \ti to
fill this gap.

\section{Dark data-reduction}\label{s-data}
The data-reduction of dark measurements consists in deglitching, and averaging
(or taking the median) of the stabilized frames. 

\subsection{Deglitching}\label{ss-deg}
Deglitching is always a very complicated operation for ISOCAM observations.
Several deglitching methods are used for the dark 
measurements\footnote{The reason why
different methods of data-reductions have been adopted for 
different dark measurements is historical and we will not discuss it here.},
but typically
the most used are {\em TCOR, TEMP,} and {\em MM,} all methods implemented
in the {\em CAM Interactive Analysis, CIA}\footnote{CIA is a joint
development by the ESA Astrophysics Division and the ISOCAM Consortium.
The ISOCAM Consortium is led by the ISOCAM PI, C. Cesarsky, Direction des
Sciences de la Mati\`ere, C.E.A., France. Contributing ISOCAM Consortium
institutes are Service d'Astrophysique (SAp, Saclay, France) and Institut
d'Astrophysique Spatiale (IAS, Orsay, France) and Infrared Processing and
Analysis Center (IPAC, Pasadena, U.S.A.).}
software package (Ott et al. 1996, Delaney 1998). 

The {\em TEMP} method of deglitching has been used in the early times
of the ISO mission for the data-reduction of \cad, while the {\em MM} method
has been used in the most recent times. The two methods provide essentially
identical results. More specifically, we can quantify the difference as
follows. After deglitching the same set of dark frames with the two methods,
we subtracted one of the resulting dark image from the other; the residuals
of this subtraction quantify the importance of using two different methods
of deglitching. The average residual difference (over all pixels) is
less than 0.005 ADU,
with a root-mean-square (RMS hereafter) of 0.04 ADU and the
maximum difference is 0.16 ADU. 

On the other hand, the {\em TCOR} method has been used for the deglitching
of \hod~ (Gallais \& Boulade 1998),
and it does give slightly different results from the previous two methods.
The final dark images as obtained from the same data-set but with either
the {\em TCOR} deglitching method, or the {\em MM} one, differ by, on average,
0.07 ADU, with an RMS of 0.12 and a peak difference of 0.34 ADU. These
differences arise from the lower deglitching efficiency of the {\em TCOR}
method. Since faint glitches are not detected by
the {\em TCOR} method, yet they are detected by the 
{\em TEMP} or the {\em MM} ones, and glitches always contribute a spurious
positive signal when the detector is at dark, 
the resulting dark level after deglitching is always
sistematically higher when the {\em TCOR} method is used.

Luckily, the \hod~ are only used for establishing a linear trend over
long timescales (see \S~\ref{s-long}) and the good statistics (\hod~
are measured every day) compensates for the additional noise induced
by an imperfect deglitching. The overall positive offset of the mean
level of \hod~ is not a problem, as it can be corrected using the
\cad~ and/or the \pad~ (see \S~\ref{s-short}).

\subsection{Stabilization}\label{ss-sta}
\begin{figure}
\centerline{\psfig{figure=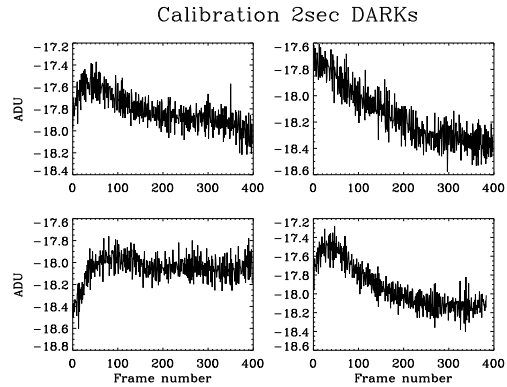,width=8cm,angle=90}}
\caption[ ]{\footnotesize{The mean dark current level vs. frame number in four
\cad~ measurements at \ti=2~sec. Note the slow stabilization drifts.}}
\label{f-cdstab}
\end{figure}

As the ISOCAM detector has a slow response to flux variations, it takes
some time for an ISOCAM pixel to level off to the dark level from a
pre-existing condition of illumination. Often, even in very long dark
measurements, stabilization is not achieved. As an example of different
behaviours, we show in Fig.~\ref{f-cdstab} the mean dark level vs. the
frame number in four \cad~ at \ti=2~sec. After an initial increase,
due to the change of integration time from a previous dark
measurement, the dark level slowly decreases to its stabilized value.
It is clear that only the last frames must be used to form the final
(stabilized) image.

\begin{figure}
\centerline{\psfig{figure=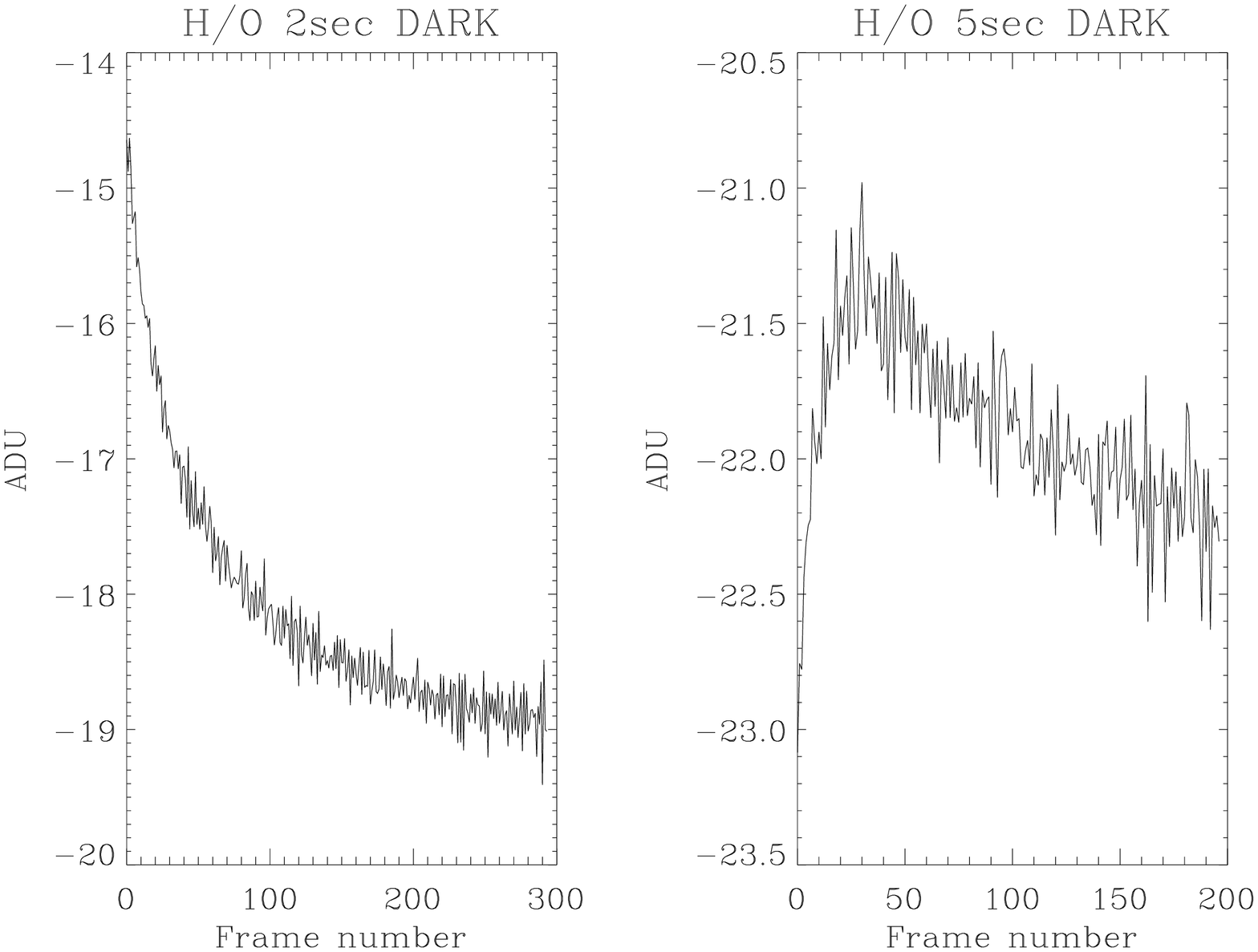,width=8cm,angle=90}}
\caption[ ]{\footnotesize{The mean dark current level vs. frame number in two \hod~
measurements at \ti=2 and 5~sec. Note the slow stabilization drifts.}}
\label{f-hdstab}
\end{figure}

As an additional example, we show in Fig.~\ref{f-hdstab} the \hod~
measured at 2 and 5~sec \ti, i.e. the mean current level measured in
the ISOCAM/LW detector vs. the frame number. In the \ti=2~sec dark
measurement, one can see the initial decrease due to the fact that ISOCAM
was open before the dark measurement; in the \ti=5~sec dark
measurement, one can see the current change due to the change in \ti,
followed by another long period of stabilization to the level of the
dark current at \ti=5~sec.

\begin{figure}
\centerline{\psfig{figure=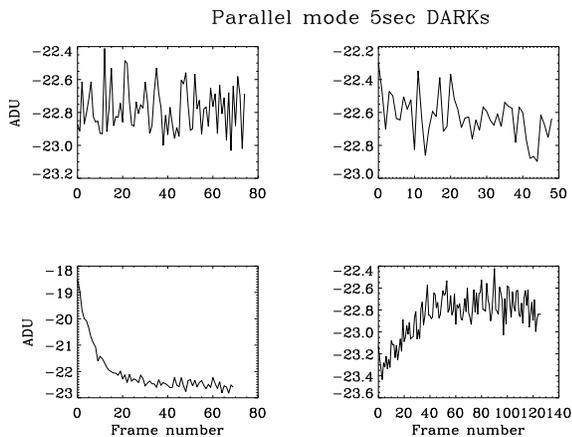,width=8cm,angle=90}}
\caption[ ]{\footnotesize{The mean dark current level vs. frame number in four \pad~
measurements at \ti=5~sec. Note the stabilization drift to the
darkness level, after the illumination
of the detector during the handover (bottom-left panel).}}
\label{f-pdstab}
\end{figure}

Finally, we give in Fig.~\ref{f-pdstab} four examples of \pad.  As
before, we plot the mean current level vs. the frame number to show
the stabilization trend. Generally, the fact that the ISOCAM detector is
kept in the same dark configuration during the whole revolution
(except for the handover interval) means that stabilization should not
be a problem here.  However, after handover there is a stabilization
drift (this is shown in the lower-left panel of Fig.~\ref{f-pdstab}),
and additional drifts (of much lower importance, though) arise because
of the change in the ISOCAM detector temperature, which is affected by
the operations of other ISO instruments (see \S~\ref{s-temp}).

The criteria for choosing the stabilized frames may be slightly different
for the three kinds of measurements. In particular, the choice of the
stabilized frames of the \cad~ cannot be blind, as these measurements
are not always done in the same conditions of previous illumination of the ISOCAM
detector (e.g., the dark measurement could be done after the
observation of a calibrating star, or after a parallel mode configuration);
depending on what was the previous ISOCAM target, the stabilization trend can
be quite different. On the other hand, \pad~ are always very close to
stabilization, so that, after excluding the frames measured just after the
handover, all frames can be used to build the dark image, by taking their
median. Finally, since the measurements of \hod~ are always done in the same
way, the data-reduction procedure is automatic, and only the last 10 frames 
are averaged to give the final dark image (see Gallais \& Boulade 1998).
This choice can
produce a different mean dark current level for \hod~ as compared to \cad~ and
\pad. However, this offset can be (and is) corrected (see \S~\ref{s-short}).

\section{Long-term drift}\label{s-long}
\begin{figure}
\centerline{\psfig{figure=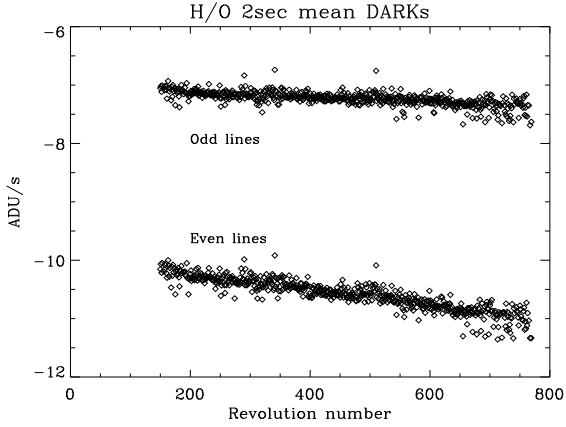,width=8cm,angle=90}}
\caption[ ]{\footnotesize{The mean values of the odd and even line pixel
dark-currents as measured during the handover, vs. the revolution
number, at \ti=2~sec.}}
\label{f-long2}
\end{figure}

\begin{figure}
\centerline{\psfig{figure=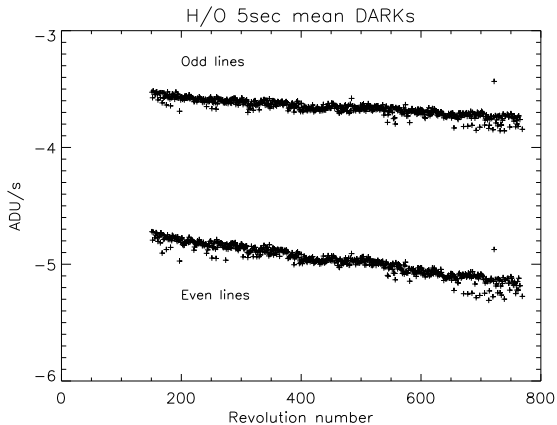,width=8cm,angle=90}}
\caption[ ]{\footnotesize{The mean values of the odd and even line pixel
dark-currents as measured during the handover, vs. the revolution
number, at \ti=5~sec.}}
\label{f-long5}
\end{figure}

The \hod~ provide an excellent constraint on the long-term drift of
the LW dark current. The mean value of the \ti=2 and 5~sec \hod~ are
shown in Fig.~\ref{f-long2} and Fig.~\ref{f-long5}, respectively, from
revolution 150 to 829 (\hod~ started to be measured routinely since
revolution 150).  There is a clear decreasing trend of the dark
current, approximately linear with the revolution number. However, the
trend is not the same over the whole array, as can be seen in the
figures, where the mean values of the odd- and even-line pixels are
shown separately. A proper characterization of this trend must
therefore take into account individual pixels.

\begin{figure}
\centerline{\psfig{figure=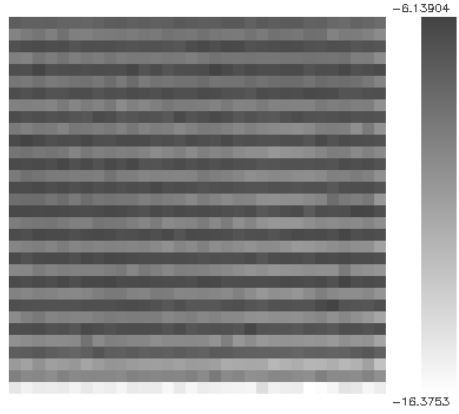,width=10cm}}
\caption[ ]{\footnotesize{The ``zero-revolution'' dark frame
as obtained from the intercepts of the linear fits to each
pixel long-term drift, at \ti=2~sec. The grey scale, on the right side,
is in ADU/s. }}
\label{f-long_inter}
\end{figure}

\begin{figure}
\centerline{\psfig{figure=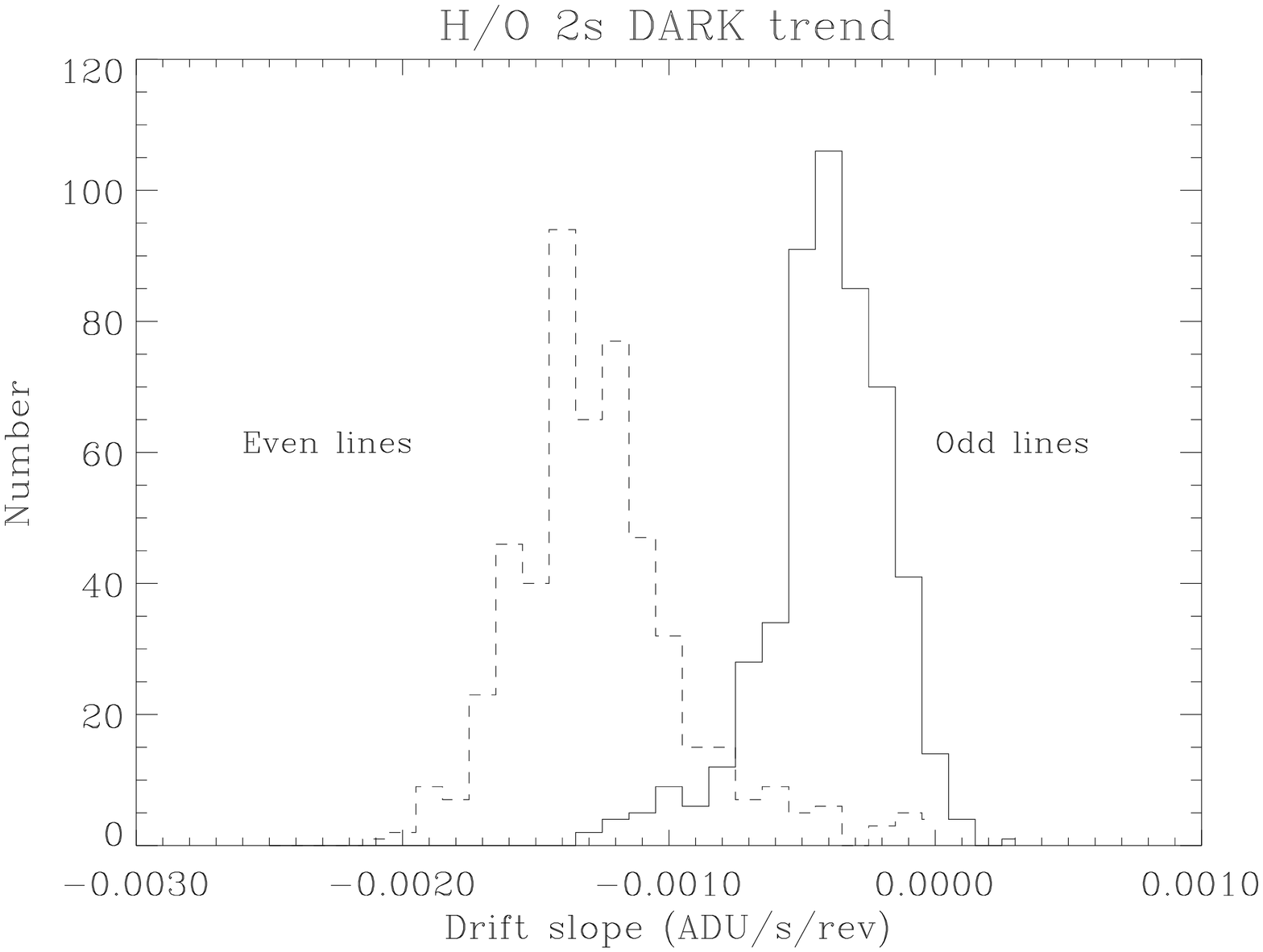,width=8cm,angle=90}}
\caption[ ]{\footnotesize{The distribution of the long-term drift
slopes, in ADU/s/rev,
of the individual pixel dark-currents, at \ti=2~sec, obtained
by a straight line fitting of the observed value of the dark current of each
pixel vs. revolution number.}}
\label{f-hlong2}
\end{figure}

\begin{figure}
\centerline{\psfig{figure=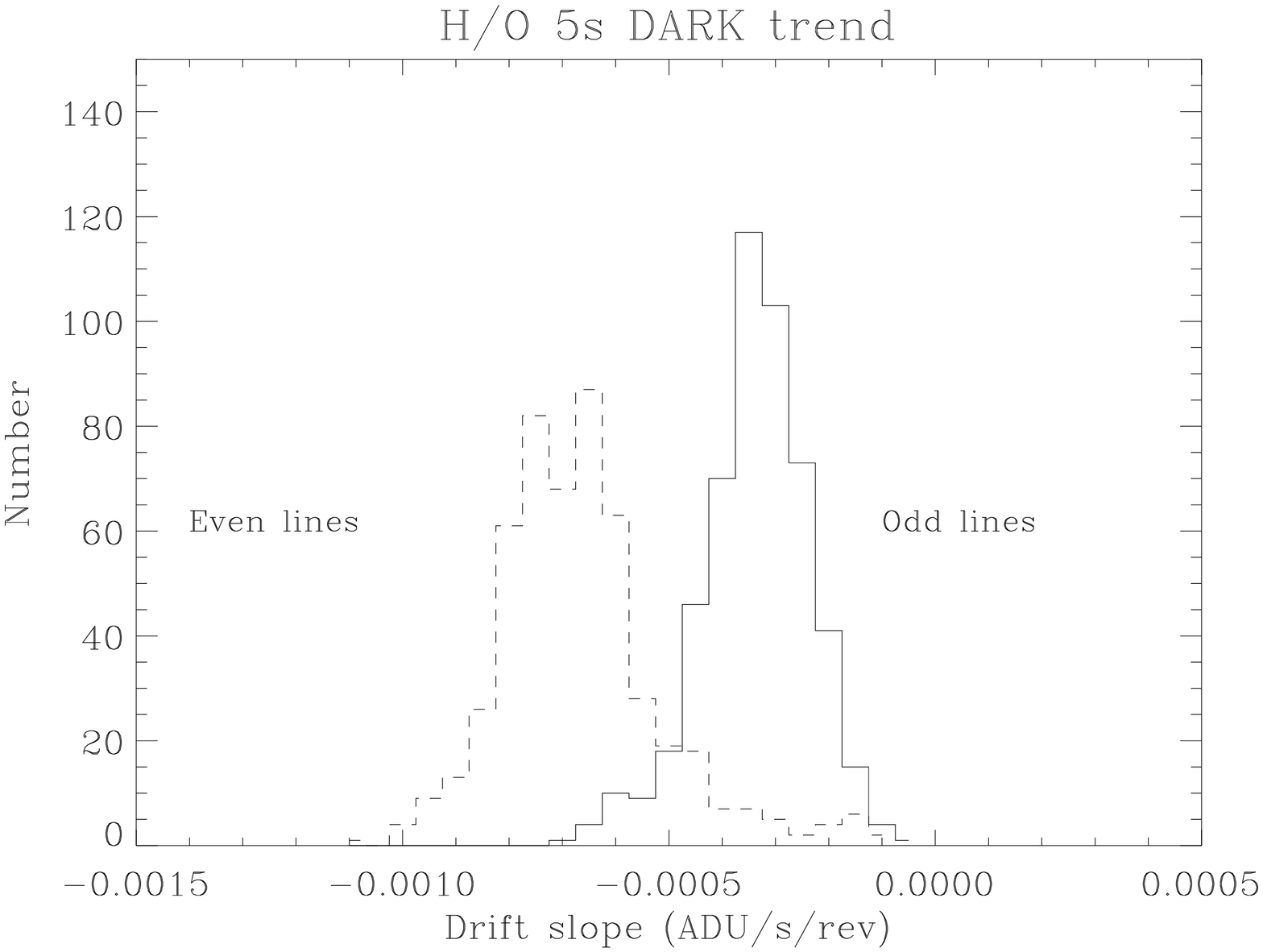,width=8cm,angle=90}}
\caption[ ]{\footnotesize{The distribution of the long-term drift
slopes, in ADU/s/rev,
of the individual pixel dark-currents, at \ti=5~sec, obtained
by a straight line fitting of the observed value of the dark current of each
pixel vs. revolution number.}}
\label{f-hlong5}
\end{figure}

For each pixel, we made a linear fit of the dark current drift with 
revolution. The fitted values of the intercepts define a ``zero-revolution''
dark frame which we show in Fig.~\ref{f-long_inter}.
The histograms of the individual pixel
long-term drift slopes are shown in Fig.~\ref{f-hlong2} and Fig.~\ref{f-hlong5}.
It can be seen that the drift is (on average) more negative for even-line 
pixels than for odd-line pixels, for both \ti, and ranges from
$\sim -0.2$~ADU/s, per 100 revolutions to $\sim 0.00$.
In order to understand how much of the difference among the individual pixel 
gradients is real, and how much is due to fitting uncertainties, we
repeated the regression analysis separately for two subsamples of equal sizes
extracted from the whole sample (i.e. by selecting only odd number revolutions 
in one subsample, and even number revolutions in another). We found that the 
average uncertainty on the individual pixel gradients is only 
$\sim 0.004$ ADU/s per 100 revolutions, much smaller than the 
pixel-to-pixel variations, which then must be real.
As a consequence, applying the same drift correction to all pixels cannot work.

The accuracy of this linear-fit characterization of the dark long-term drift 
can be estimated by computing the RMS variation of the
dark current values of all pixels through the mission. 
The average per-pixel RMS of all \hod~ is $\simeq 0.5$~ADU for both \ti, 
if we do not apply any correction for the long-term drift. 
After subtracting from the observed dark current values the fitted ones, the
average per-pixel RMS of all \hod~ is reduced to $\simeq 0.25$~ADU.

Other \ti's are not well covered in the handover measurements. In particular,
the handover measurements at \ti=0.28 and 10~sec started in revolution
764, i.e. only 4 months before the end of the ISO mission.
This time coverage is too limited to allow a proper characterization
of the trend throughout the whole ISO mission. The situation concerning the
\ti=20~sec LW darks is even worse, no measurement being done in handover
for this integration time. However, it was possible to make use of the 
available \cad~ and \hod~ to constrain the global trend, by properly
scaling the linear-fits made on the 2 and 5~sec measurements of \hod. 

We first checked that there was a global consistency among the
long-term trends of the \hod~ measured at \ti=2 and 5~sec. Indeed we
found that by applying a suitable scaling factor one could deduce the
trend at one \ti from the other. Then we checked that the 2~sec
long-term trend was consistent with the available data for \cad~ at
\ti=0.28~sec, and, correspondingly, that the 5~sec long-term trend was
consistent with the available data for \hod~ and \cad~ at \ti=10 and
20~sec.  A global consistency was found.  The scaling was then
computed per pixel by averaging a set of dark measurements done at
different \ti's, during the same epoch of the ISO mission.
Specifically, the dark currents at 0.28~sec and 10~sec were computed
by averaging the handover measurements of revolutions 764 to 828. In
order to obtain the scaling factors, they were divided by the dark
currents at 2~sec and, respectively, 5~sec, computed by averaging the
handover measurements of revolutions 765 to 829. The lack of handover
measurements for the \ti=20~sec dark made this approach impossible, so
we simply used a global average of all available measurements to
perform the scaling.

The value of the dark current corrected for this long-term drift,
$D_R$, is obtained from the observed value through the relation:
\begin{equation}
D_R=D_O-(r_0 + r_1 \times rev)
\label{e-darkr}
\end{equation}
where $D_O$ is the observed value of the dark current, in ADU,
$rev$ is the revolution number, and $r_0$ and $r_1$ are the intercept
(in ADU) and slope (in ADU/rev) obtained from the linear fit to the
long-term trend. The model parameters are different for different
pixels and different \ti.  In Table~1 we list the
average values of these parameters, $<r_0>$ (in ADU) and $<r_1>$ (in
ADU/rev), separately for odd- and even-line pixels, at all \ti.

\begin{table}[htb]
\begin{center}
\caption{Results of the linear fits to the long-term trends}
\begin{tabular}{rrrrr}
\hline 
\ti&
Odd-line pixels &
Even-line pixels &
Odd-line pixels &
Even-line pixels \\
& mean intercept & mean intercept & mean slope & mean slope \\
(seconds) & $<r_0>$ (ADU) & $<r_0>$ (ADU) & $<r_1>$ (ADU/rev) & 
$<r_1>$ (ADU/rev) \\
\hline
0.28 & -7.8 & -12.8 & -0.0004 & -0.0015 \\ 
2 &  -14.8 & -21.0 & -0.0008 & -0.0025 \\
5 &  -17.4 & -23.4 & -0.0016 & -0.0032 \\
10 & -19.3 &  -25.5 & -0.0017 & -0.0033 \\
20 & -20.2 &  -26.8 & -0.0020 & -0.0036 \\
\hline
\end{tabular}
\end{center}
\end{table}

\section{The temperature dependence}\label{s-temp}
\begin{figure}
\centerline{\psfig{figure=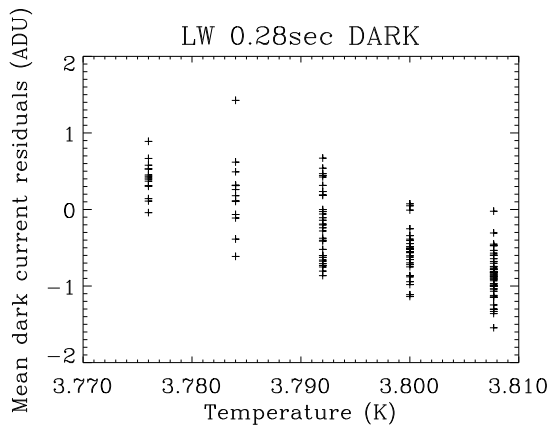,width=8cm}}
\caption[ ]{\footnotesize{The correlation between the long-term
drift-corrected mean value of the LW dark current at \ti=0.28~sec, as
observed in several calibration revolutions, and the ISOCAM detector
temperature, in K.}}
\label{f-temp}
\end{figure}

Once the dark measurements are corrected for the revolution drift, it
is possible to show the existence of a relation between the mean dark
current and the temperature of the ISOCAM detector.  This trend is
shown in Fig.~\ref{f-temp} for the \ti=0.28~sec \cad, after correction
for the long-term drift. At the current status of understanding, we
can assume that the trend is the same for all pixels and all \ti's.
The dark current corrected for this temperature dependence, $D_{R,T}$,
is obtained through the following relation: 
\begin{equation}
D_{R,T}=D_R-66.119 + 17.467 \times T 
\label{e-darkt}
\end{equation}
where $T$ is the temperature in K, and $D_R$ is the
value of the dark current, in ADU, corrected for the long-term drift
(see \S~\ref{s-long}).

\section{Short-term drifts}\label{s-short}
The short-term drift of the dark current, repeating at each
revolution, from the activation of the instrument to its deactivation,
was first discovered in the \cad. At variance with the \hod, the \cad~
are {\em not} done at approximately the same time within a revolution.
In some cases, several measurements were done during the {\em same}
revolution at different times. It is then possible to look for the
existence of a drift of the dark current with the time elapsed since
the activation of the ISOCAM instrument in any revolution (the ``time
since activation\footnote{The choice of the reference time within each
revolution, for what concerns the dark short-term drift, is not
unique. Instead of choosing the activation time, the perigee passage
of ISO could be chosen as well. In most revolutions, the time of
activation is strictly related to the time of perigee passage, but in
some it departs from the usual relation. In these, the short-term
drift correction we have derived -- described in this section --
should be poorer than average, if the origin of time is not related to
the activation, but, rather, to the perigee passage. This remains to
be tested.}'', \tsa hereafter).

\begin{figure}
\centerline{\psfig{figure=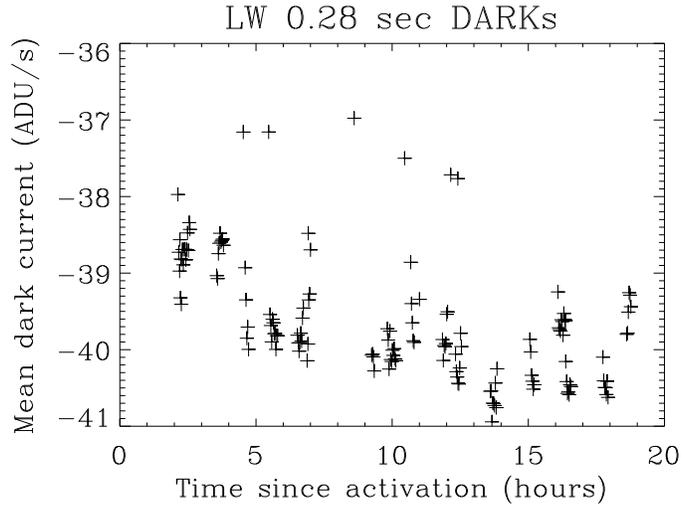,width=10cm}}
\caption[ ]{\footnotesize{The mean value of the LW detector dark-current,
measured at \ti=0.28~sec in several calibration revolutions, vs. the \tsa.}}
\label{f-lwtsa}
\end{figure}

In Fig.~\ref{f-lwtsa} we plot the mean dark current value of the LW
detector, vs. the \tsa, for several \ti=0.28~sec \cad. There is a very
clear correlation between the two variables. Since the \cad~ are taken
in different revolutions throughout the mission, the real trend may
however be partly masked by the long-term drift with revolution.

\begin{figure}
\centerline{\psfig{figure=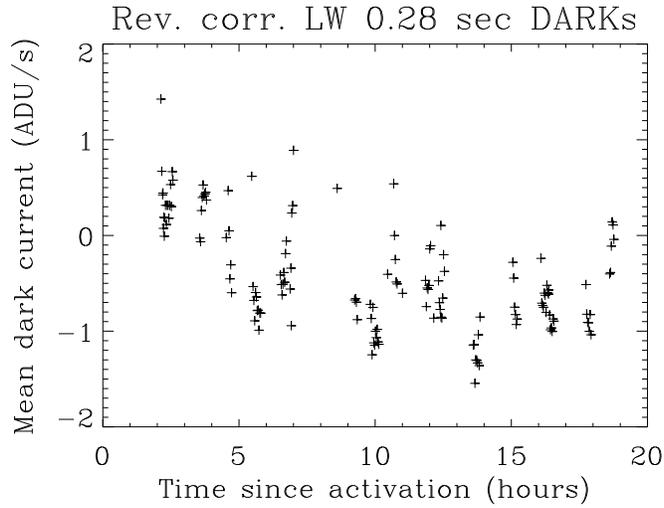,width=10cm}}
\caption[ ]{\footnotesize{The residual mean dark current of the LW
detector, after subtraction of the long-term trend model, and the
temperature dependence model, measured at \ti=0.28~sec in several
calibration revolutions, vs. the \tsa.}}
\label{f-lwtsac}
\end{figure}

\begin{figure}
\centerline{\psfig{figure=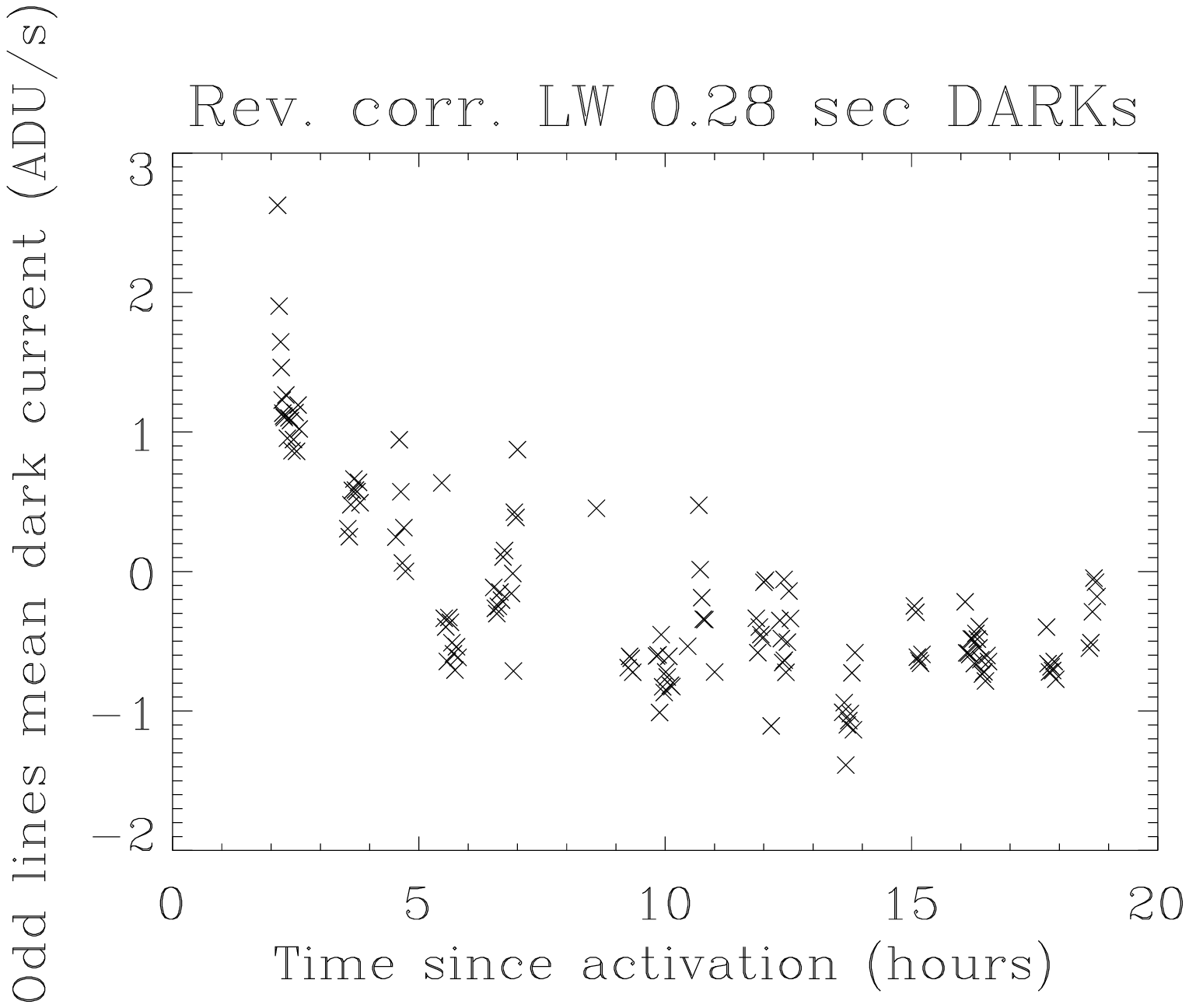,width=10cm}}
\caption[ ]{\footnotesize{The residual mean dark current of the
odd-line pixels of the LW detector, after subtraction of the long-term
trend model, and the temperature dependence model, measured at \ti=0.28~sec in
several calibration revolutions, vs. the \tsa.}}
\label{f-lwtsaco}
\end{figure}

\begin{figure}
\centerline{\psfig{figure=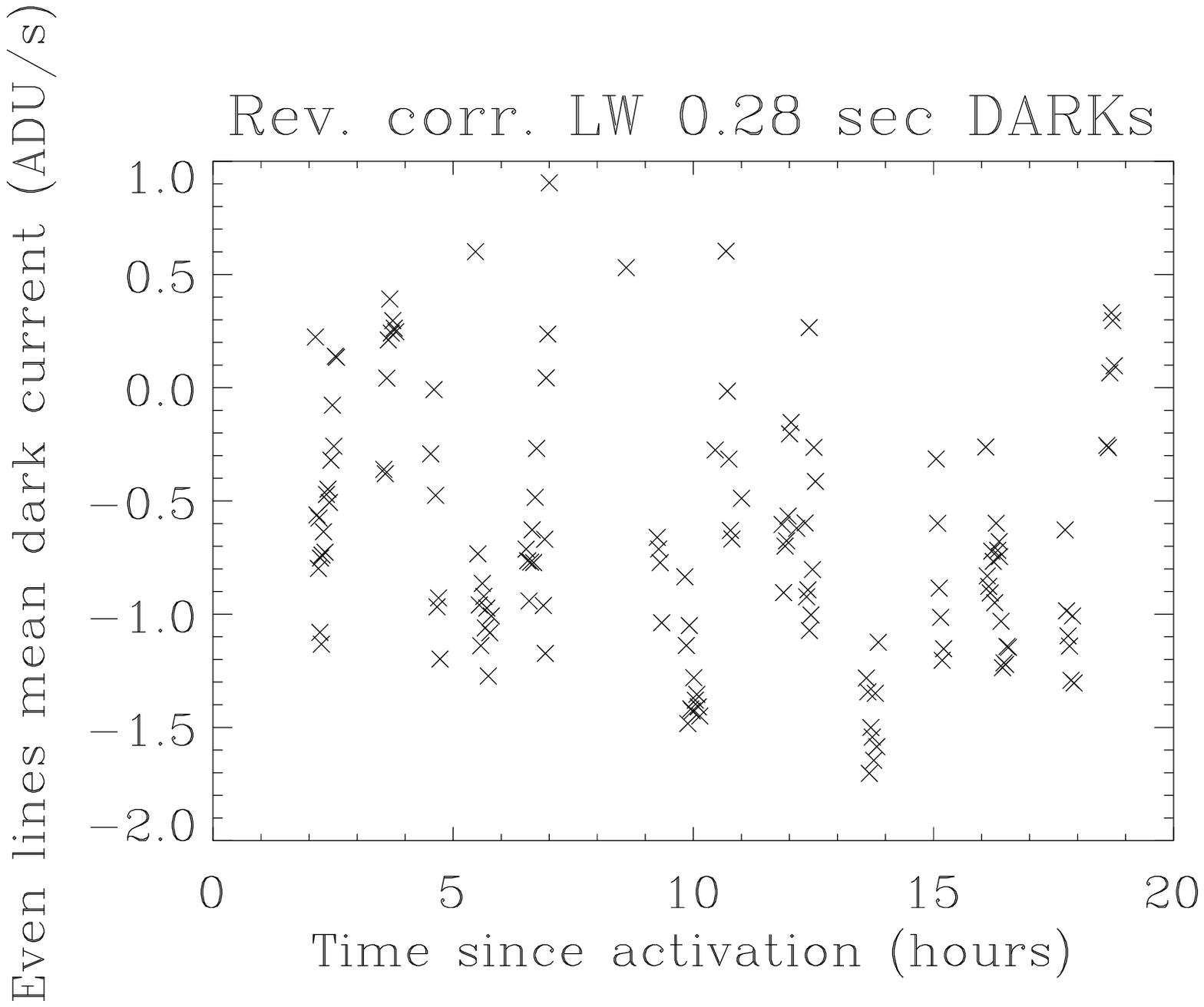,width=10cm}}
\caption[ ]{\footnotesize{The residual mean dark current of the
even-line pixels of the LW detector, after subtraction of the
long-term trend model, and the temperature dependence model, measured at
\ti=0.28~sec in several calibration revolutions, vs. the \tsa.}}
\label{f-lwtsace}
\end{figure}

\begin{figure}
\centerline{\psfig{figure=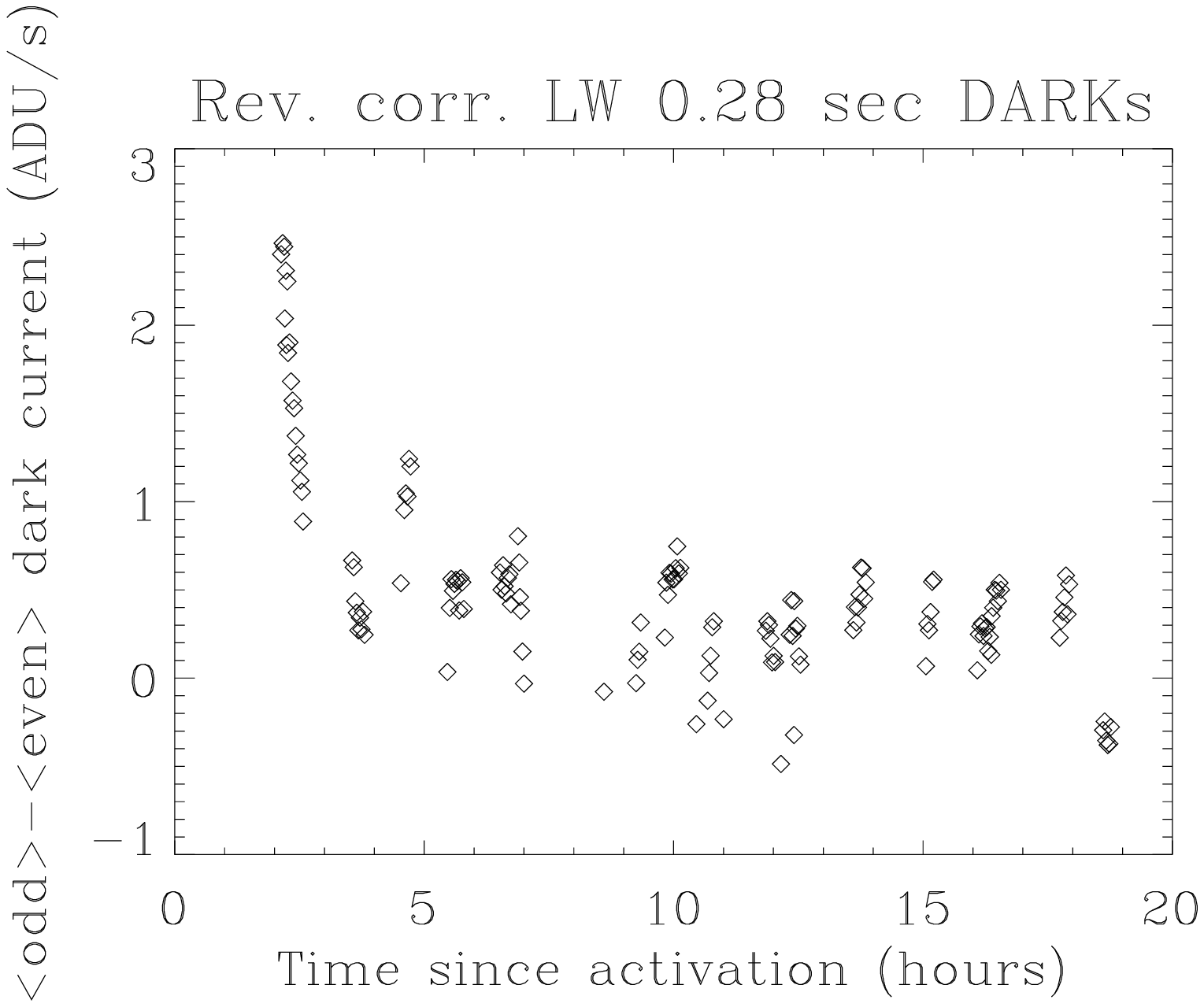,width=10cm}}
\caption[ ]{\footnotesize{The difference of the mean dark currents of
the odd- and even-line pixels of the LW detector, after correction for
the long-term trend, and the temperature dependence, measured at
\ti=0.28~sec in several calibration revolutions, vs. the \tsa.}}
\label{f-lwtsacoe}
\end{figure}

We corrected the LW \cad~ for the long-term trend. We subtracted to
the observed value of the dark current, the linear fit to the
long-term trend of the \hod~ (see \S~\ref{s-long}, eq.~\ref{e-darkr}
and Table~1), on a per pixel basis. We then subtracted
the (less strong) temperature dependence model (see \S~\ref{s-temp},
eq.~\ref{e-darkt}). On the resulting residual dark current, the
short-term trend with \tsa is more evident (see Fig.~\ref{f-lwtsac}).
Similarly to the long-term drift, also in this case the trend is
different for odd-line and even-line pixels. A proper characterization
of the trend is infact only possible on a per pixel basis.  As an
example, we plot in Fig.~\ref{f-lwtsaco} and Fig.~\ref{f-lwtsace} the
trend of the mean dark current (at \ti=0.28 sec) with \tsa for the odd and,
respectively, the even line pixels, and in Fig.~\ref{f-lwtsacoe} the
trend of the difference of the two. 

\begin{figure}
\centerline{\psfig{figure=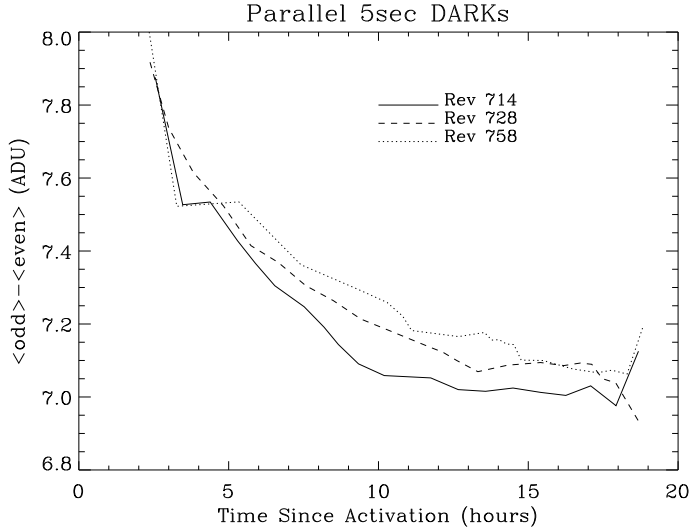,width=10cm,angle=90}}
\caption[ ]{\footnotesize{The (median-filtered) difference of the mean
dark currents of the odd and even line pixels of the LW detector,
measured in parallel mode at \ti=5~sec in three revolutions, vs. the
\tsa.  No correction for the long-term trend has been applied to the
data, so that a displacement is visible from revolution to
revolution.}}
\label{f-shortpar}
\end{figure}

The \pad~ also show the short-term drift very clearly. This is
illustrated in Fig.~\ref{f-shortpar}, where the trend of the
difference in the mean values of the dark current for odd and even
pixels is shown, vs. \tsa. The measurements were taken at \ti=5~sec in
three different revolutions. In this case, the raw observed values of
the dark current have been used, without correcting for the long-term
trend and the temperature dependence. Infact, the long-term drift is
also visible in the figure, showing up as a small vertical shift
in the short-term drift curve, from one revolution to another.

The short-term drift appears to be as important as the long-term drift
for the purpose of a correct calibration of the LW dark current behaviour.
We made use of the \pad~ for characterizing this trend for all \ti except
for \ti=0.28~sec, for which we used the \cad~ (\pad~ are not available for
this \ti, see \S~\ref{s-intro}). Before making a fit of the dark current
trends with \tsa for each pixel, we corrected the observed darks for 
the long-term trend and the temperature dependence, as
described in \S~\ref{s-long} (see eq.~\ref{e-darkr}) and \S~\ref{s-temp}
(see eq.~\ref{e-darkt}).

At variance with the long-term trend, a linear fit is not a proper
characterization of the short-term drift for all pixels. For some
pixels a quadratic or even a cubic fit was needed. More specifically, a
cubic fit was needed for properly describing the trend of the
\ti=2~sec dark current of even line pixels. Quadratic fits were needed
for characterizing the short-term drifts of the dark currents of odd
line pixels at \ti=0.28, 2 and 10~sec, as well as that of the even
line pixels at \ti=5 and 20~sec. Finally, linear fits were adopted in
all the other cases. 

The dark current corrected for this short-term drift is obtained through:
\begin{equation}
D_{R,T,S}=D_{R,T}-(s_0 + s_1 \times tsa + s_2 \times tsa^2
+ s_3 \times tsa^3) 
\label{e-darks}
\end{equation}
where $D_{R,T}$ is the dark current corrected for the long-term drift
and for the temperature dependence, in ADU, and $s_0, s_1, s_2, s_3$ are
the intercept, and the linear, quadratic and cubic coefficients of the
fit of the short-term trend. The coefficients are different for different
pixels and different \ti's. The average values of these coefficients are
given in Table~2 for all \ti's.

\begin{table}[htb]
\begin{center}
\caption{Results of the fits to the short-term trends}
\begin{tabular}{rrrrr}
\hline 
Integration time &
Intercept & Linear coefficient & Quadratic coefficient & Cubic coefficient \\
(seconds) & 
$<s_0>$ (ADU) & $<s_1>$ (ADU/hour) & 
$<s_2>$ (ADU/hour$^2$) & $<s_3>$ (ADU/hour$^3$) \\
\hline
0.28  &  0.2 & -0.046 &  0.0016 & 0.00000 \\ 
   2  & -0.3 &  0.017 & -0.0092 & 0.00025 \\
   5  & -0.6 &  0.026 & -0.0018 & 0.00000 \\ 
  10  & -0.4 & -0.030 &  0.0010 & 0.00000 \\ 
  20  &  1.0 & -0.100 &  0.0020 & 0.00000 \\ 
\hline
\end{tabular}
\end{center}
\end{table}

Note that the intercept of the fitting lines (or the zero coefficient
of the polynomial fitting) is in general different from zero. This is
because the correction applied for the long-term trend was based on
\hod~ which have a slightly higher level than other dark measurements
because of the different deglitching routine used in the
data-reduction, and because of the poorer stabilization of the
detector to the darkness (see \S~\ref{ss-deg} and \S~\ref{ss-sta}).
The application of the short-term drift correction also corrects for
this small systematic offset.

\begin{figure}
\centerline{\psfig{figure=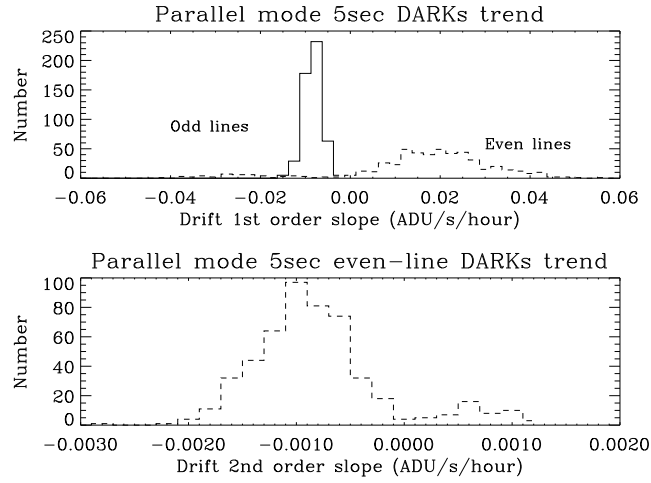,width=9cm,angle=90}}
\caption[ ]{\footnotesize{The distribution of the short-term drifts of the
dark currents of individual pixel, at \ti=5~sec; note that the fits of
the even-line pixel trends are quadratic, while those of the odd-line pixels
are linear.}}
\label{f-hshort5}
\end{figure}

The distributions of the linear and quadratic terms of the fitting to
the short-term drifts of the individual pixel dark currents are shown
in Fig.~\ref{f-hshort5} for \ti=5~sec.  The odd line pixels have a
decreasing trend with \tsa, while the even line pixels have an
increasing trend. This property is characteristic of the dark current
short-term drifts of all \ti's.

\section{The accuracy of the calibration}\label{s-model}
As discussed in the previous sections, our calibration of the dark
current for the LW detector takes into account both a long-term and a
short-term drift (with a timescale of days, and respectively, hours),
as well as a (minor) variation with the focal-plane temperature.  In
practice, we have built a phenomenological model for the dark current
that takes into account these dependencies. It can be expressed by the
following relation:
\begin{equation}
D_{model} = r_0 + r_1 \times rev +
s_0 + s_1 \times tsa + s_2 \times tsa^2
+ s_3 \times tsa^3 + 66.119 - 17.467 \times T 
\label{e-darkm}
\end{equation}
where the algebraic coefficients depend on the pixel and the \ti,
$rev$ is the revolution number, \tsa the time since activation in
hours, and $T$ the temperature of the ISOCAM detector in K.

\begin{figure}
\centerline{\psfig{figure=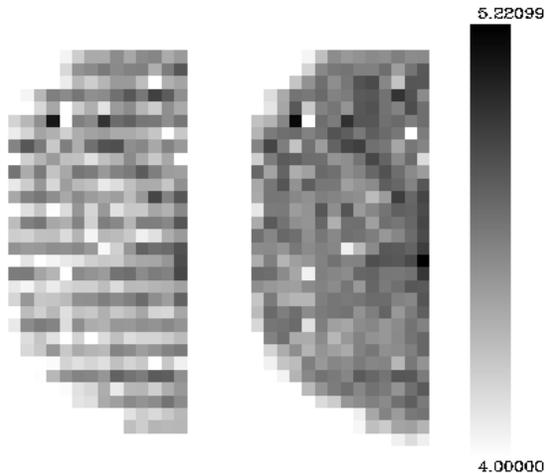,width=9cm}}
\caption[ ]{\footnotesize{The calibration observation of the star HIC85317 
done in rev.419, at 6.0\asec PFOV, and \ti=2~sec, with the LW6 filter,
reduced using an average dark (left), and the model
dark described in this paper (right). Only half of the array
is shown in both cases (the star itself is not visible, since it is located on the
other half of the array). Flat-fielding was not applied, which
explains the vignetting effect at the edges.}}
\label{f-hic85317dark}
\end{figure}

The precision of our calibration can be evaluated by computing the residuals
of the subtraction of our model dark (eq.~\ref{e-darkm})
from individual \cad. From all
dark-subtracted dark measurements at a given \ti, we compute the
per-pixel RMS. The median RMS is typically $\simeq 0.25$ ADU, similar
to the one found for the \hod~ without any temperature and \tsa-drift
correction, but after the long-term trend correction (see
\S~\ref{s-long}). This is not unexpected, since the handover darks are
all done at roughly the same \tsa, so there is no need for correcting
them for the \tsa-drift, and the temperature dependence is generally
of minor importance as compared to the revolution and \tsa drifts.

We can compare this time- and temperature-dependent dark current
correction with that obtained by a simple average of several
dark-current measurements, i.e.  the ``calibration dark" that has been
the standard correction applied for all ISOCAM data-reduction until
recently. Using the traditional average dark current correction would
increase the median per-pixel RMS of the residuals to $\simeq 0.5$
ADU, i.e. twice the value we obtained using the model dark.

In practice, using our dark model in the data-reduction of ISOCAM
observations, greatly improves the quality of the images as compared
to the traditional data-reduction (when a simple average of several
dark current observations was used).  This is illustrated in the
following examples.

\begin{figure}
\centerline{\psfig{figure=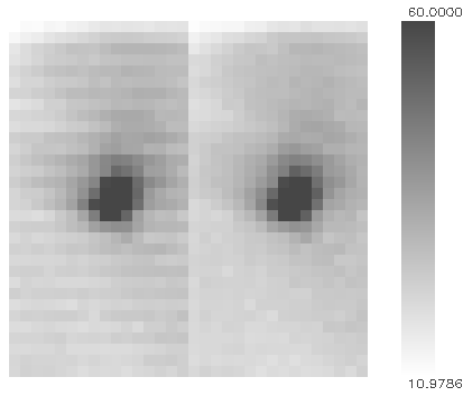,width=10cm}}
\caption[ ]{\footnotesize{The image of a comet, observed in rev.341,
with \ti=0.28~sec, with the LW9 filter. The observed data have been
reduced using a simple average of many darks (left), and using the
model dark described in this paper (right). Only the central half of
the array is shown in both cases. The images are shown before
flat-fielding.}}
\label{f-comet}
\end{figure}

\begin{figure}
\centerline{\psfig{figure=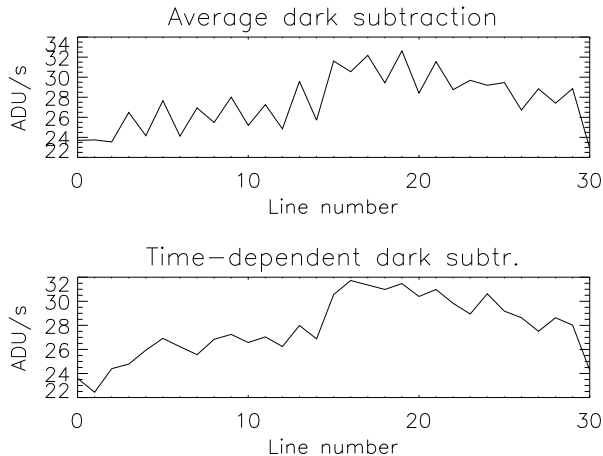,width=9cm}}
\caption[ ]{\footnotesize{The flux along a single column of the comet
image shown in Fig.~\ref{f-comet}. Upper panel: after subtraction of
an average calibration dark. Lower panel: after subtraction of the model
dark described in this paper.}}
\label{f-comet1}
\end{figure}

In Fig.~\ref{f-hic85317dark} we provide an example of a specific
calibration observation of a star, at \ti=2~sec.  The data-reduction
was done first by using an average dark correction, and then using our
model dark correction.  The odd/even line pattern is only evident in
the image reduced with the average dark, indicating a poor dark
correction.

Another example is given in Fig.~\ref{f-comet}, where we show the
image of a comet observed at \ti=0.28~sec. Again, the odd/even line
pattern is evident when we apply the average dark subtraction but not
anymore when our dark model is used.  To have a quantitative feeling
of the accuracy of the dark correction, we plot in Fig.~\ref{f-comet1}
the fluxes measured along the pixels of a single column of the ISOCAM
detector, so to emphasize the line pattern. The residual is $\sim
2$~ADU/s (i.e.  0.6~ADU, since \ti=0.28~sec) when the average dark
subtraction is applied, while it is less than 1~ADU/s ($\sim 0.2$~ADU)
when the dark model subtraction is applied,
consistently with the estimates given above.

\begin{figure}
\centerline{\psfig{figure=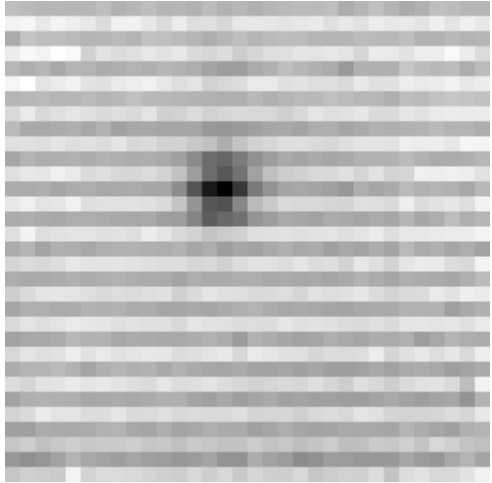,width=9cm}}
\caption[ ]{\footnotesize{The final reduced image of a 100~Mpc distant
lenticular galaxy, observed in rev.617, in the LW2 filter, with
1.5\asec~ pixel-field-of-view, at \ti=2~sec. The dark used in the
data-reduction is an average of several \cad.}}
\label{f-gmpcg}
\end{figure}

\begin{figure}
\centerline{\psfig{figure=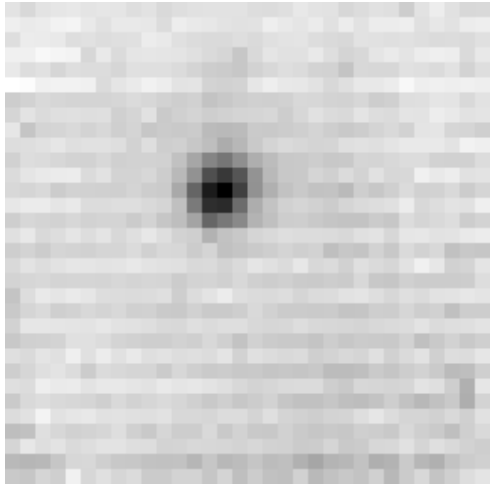,width=9cm}}
\caption[ ]{\footnotesize{The final reduced image of the same observation
of Fig.~\ref{f-gmpcg}, with the dark model used in the data-reduction,
instead of an average calibration dark.}}
\label{f-gmptime}
\end{figure}

As a final example, we show in Fig.~\ref{f-gmpcg} and
Fig.~\ref{f-gmptime} two images of a galaxy at 100~Mpc distance,
observed witt a pixel-field-of-view of 1.5\asec~ in the LW2 filter, at
\ti=2~sec. As before, the two images have been obtained using two kind
of dark corrections in the data-reduction: an average of several \cad~
in Fig.~\ref{f-gmpcg} and our dark model in Fig.~\ref{f-gmptime}. The
dark current subtraction is not perfect even when using our dark
model, but it is anyway much better than the one obtained with a
simple average dark.

\section{Conclusions}\label{s-conc}
We have shown that there are several effects that contribute to modify 
the dark current of the ISOCAM/LW detector during the ISO mission. In particular,
we have identified:
\begin{itemize}
\item a {\em long-term drift,} i.e. a linear decrease of the dark current
from a revolution to another, with a typical time-scale of days; this
drift is different for different pixels and different \ti's;
\item a {\em short-term drift,} i.e. a modification of the dark current
within a single revolution, from the beginning of the revolution to the end;
this drift is in general non-linear, and is different for different pixels
and different \ti's;
\item a correlation of the dark current with the temperature of the 
ISOCAM focal-plane; to the current level of understanding, 
this correlation is unique, the same for all pixels,
and indipendent of the \ti.
\end{itemize}

We can summarize as follows the impact of each of
these effects on the dark-current:
\begin{itemize}
\item {\em long-term trend:} the drift is $\sim -2$~ADU 
from the beginning to the end of the ISO mission;
\item {\em short-term trend:} the drift within each revolution is
$\sim \pm 0.5$~ADU from the activation to the de-activation of the
ISOCAM instrument;
\item {\em temperature-dependence:} $\sim \pm 0.25$~ADU over the full range of 
temperature variation (few hundreths of a kelvin).
\end{itemize}

Correcting for all these effects is clearly important. We
have modelled them using polynomial fittings of the
available dark current data, taken under many different conditions (during
the handover, during prime-mode calibration revolutions, and with ISOCAM operating
in parallel-mode). The result is a dark-model, dependent on three variables,
the number of the revolution, the time of observation within that revolution
(counted since the instrument activation),
and the temperature of the ISOCAM detector. The dependence is different
for different pixels, and slightly different for different \ti's.
Note that the temperature-dependence, at
variance with the revolution and \tsa dependences, is assumed to be
the same for all \ti's and all pixels, although we are aware that future 
(more detailed) analyses may lead us to drop this assumption.

Once corrections are applied, the median LW dark current residual
amounts to $\simeq 0.25$ ADU, which is even better than the pre-launch
estimate of 0.3~ADU (see, e.g., Biviano 1998).

The success and simplicity of this modelisation of the ISOCAM/LW
dark-current behaviour has prompted its implementation in the CIA
package for the ISOCAM data-reduction (version 3.0; Delaney 1998) and
in the {\em Off-Line Processing} of the ISOCAM data (version 6.0; see,
e.g., Siebenmorgen et al. 1998).  As we have shown in \S~\ref{s-model}
this represents a significant improvement in the data-reduction of
ISOCAM data, as compared to the traditional approach which used a simple
average of many \cad, as the reference calibration dark.

\end{document}